\documentclass[
  aps,
  prb,
  twoside,
  twocolumn,
  showpacs,
  floatfix,
]{revtex4}
\usepackage{amsmath}
\usepackage{amssymb}
\usepackage{graphicx}
\usepackage{tabularx}
\usepackage{dcolumn}
\usepackage{longtable}
\usepackage{footnote}

\begin{document}

\newcommand{\bec}{\begin{center}}
\newcommand{\ec}{\end{center}}
\newcommand{\be}{\begin{equation}}
\newcommand{\ee}{\end{equation}}
\newcommand{\beqn}{\begin{eqnarray}}
\newcommand{\eeqn}{\end{eqnarray}}
\newcommand{\bet}{\begin{table}}
\newcommand{\ent}{\end{table}}
\newcommand{\bib}{\bibitem}

% commands for structuring
\newcommand{\sect}[1]{Sect.~\ref{#1}}
\newcommand{\fig}[1]{Fig.~\ref{#1}}
\newcommand{\Eq}[1]{Eq.~(\ref{#1})}
\newcommand{\eq}[1]{(\ref{#1})}
\newcommand{\tab}[1]{Table~\ref{#1}}

% commands for mathmode
\renewcommand{\vec}[1]{\ensuremath\boldsymbol{#1}}
\renewcommand{\epsilon}[0]{\varepsilon}

% comments at the margin
\newcommand{\cmt}[1]{\emph{\color{red}#1}%
  \marginpar{{\color{red}\bfseries $!!$}}}

%\topmargin 0.5cm
%\setlength{\textwidth}{18cm} 
%\setlength{\textheight}{25.5cm} 

%\baselineskip 4mm 
%\wideabs{

\title{
%\LARGE{The parameter fitting of the generalized Albe-Erhart type
%The parameter fitting of the 
Bond order potential for FeSi
}

%\LARGE{

\author{P. S\"ule} \email{sule@mfa.kfki.hu} 

\affiliation
  {Research Institute for Technical Physics and Material Science,\\
Konkoly Thege u. 29-33, Budapest, Hungary,sule@mfa.kfki.hu,www.mfa.kfki.hu/$\sim$sule,\\
}

\date{\today}

%}
%\maketitle

\begin{abstract}
  A new parameter set has been derived for FeSi using the Albe-Erhart-type bond order
potential (BOP) and the PONTIFIX code for fitting the parameters
on a large training set of various polymorphs.
%the most stable B20 ($\epsilon$-FeSi) and B2 (CsCl) phases together with unstable
%ones such as B1 (NaCl) and the dimer.
Ab initio calculations are also carried out to study the relative
stability of various polymorphs and to use the obtained data in the
training set.
The original BOP formalism was unable to account for the
correct energetic relationship between the B20 ($\epsilon$-FeSi) and B2 (CsCl) phases and
notoriusly slightly favors incorrectly the B2 polymorph.
In order to correct this improper behavior
the BOP potential has been extended by a Columbic term (BOP+C)
in order to account for the partial ionic character of FeSi. 
Using this potential we are able to account for the correct
phase order between the most stable B20 and B2 (CsCl) polymorphs 
when the net atomic charges are properly set.
Although this brings in a new somewhat uncertain parameter (the net charges)
one can adjust properly the BOP+C potential for specific problems.
To demonstrate this we study
under high pressure the B2 phase which becomes more stable vs. B20
as it is found experimentally and expected to be taken place in the Earth
mantle.
The obtained BOP has also been tested for the metallic and semiconducting
disilicides ($\alpha$-FeSi$_2$ and $\beta$-FeSi$_2$) and
for the Si/$\beta$-FeSi$_2$ heterostructure.
The obtained BOP, as many other BOP, overestimates the
melting point ($T_m$) of the B20 phase by $\sim 1000$ K 
if the parameters in the radial part of
the potential were obtained according to the Pauling relation (regular way).
Hence a special attention is paid 
to the $T_m$ porblem.
It has been realized that if the dimer parameters ($D_0$ and $r_0$)
are adjusted irregulary (using shorter $r_0$ and a larger $D_0$ than those of the dimer), a remarkable improvement can be reached on $T_m$
while the other properties of the potential remains nearly unaltered
(except the dimer properties).
We release a few sets of parameters in order to make a detailed comparative
test and to demonstrate that the use of the anisotropic parameter space
a significant improvement can be achieved in the melting properties with the BOP.

\pacs{68.35.-p, 79.20.Rf, 81.65.Cf, 61.82.Fk, 96.35.Gt}
%{\scriptsize {\em Keywords:} computer simulations, ion-sputtering, ion-solid interaction, molecular dynamics, Si, nanopatterning, ripples, surface physics
%}
\end{abstract}
%}

\maketitle

\section{Introduction}

 Iron-silicides have been the subject of numerous studies
in various fields of Materials Science.
Earth scientists consider ion-silicides as a main
component of the Earth-Mantle and as such could play
an important role in many processes occur in the inner region of Earth \cite{meltinner,geology}.
In particular, the stabilization of B2 (CsCl) FeSi has been studied in detail
under high pressure vs. the B20 $\epsilon$-FeSi at ambient conditions \cite{meltB2}.

 Computer simulations are limited to the  {\em ab initio} 
density functional theory level (DFT)
which does not allow the simulation of large scale systems \cite{Vocadlo}.
This is because an adequate empirical potential 
is missing in the literature for FeSi.
The main motivation of this work was to fulfill this gap.

 Contrary to the importance of FeSi as a basic material,
at best of our knowledge there is no reailable parameter set is
available for any kind of a empirical potentials including
either Buckingham-type (see e.g. ref. \cite{Buckingham}) nor bond order-type potential functions \cite{Tersoff-Brenner} which
are the basic candidates for such a binary compound.
While the previous one is a simple pair potential and does not account for
many-body effects, the Tersoff-Brenner formulation of bond order potentials (BOP)
treat 3-body effects, e.g. bond angles adequately (angular dependence), hence a more appropriate
choice for compounds with a considerable covalent characters.
Unfortunately, the BOP lacks long range effects, such as Coulomb or
van der Waals interactions.

 Recent interesting results urge the development of a
new parameter set for FeSi. These experiments on Fe-contamination driven
nanopatterning on Si \cite{rippleFe}, or recent speculations on the role of FeSi plays
in the Earth mantle \cite{geology}, on nanowires \cite{NW} or on various interface structures \cite{if}, just to mention few examples,
prove the importance of computer simulations in FeSi-related problems.
Recently published results provided evidences that
nanowires made from the semiconducting $\beta$-FeSi$_2$ are potentially applicable for spintronic nanodevices
and exhibit photoluminescence
\cite{bfesi2}.
Solar cell applications as well as the integration of nanodevices
into the Si-technology could also be feasible in the near future
using $\beta$-FeSi$_2$ nanowires \cite{bfesi2_tech}.

 Recent success in the development of the Albe-type bond order potentials (A-BOP) \cite{Albe,Erhart}
for various materials (either in covalent, metallic or ionic nature,
such as PtC \cite{Albe}, GaN \cite{GaN}, ZnO \cite{ZnO}, GaAs, WC, SiC \cite{SiC}, etc.)
provided evidence for the accuracy and effectiveness of the BOP formulation
of empirical potentials in the prediction and reproduction of various
materials properties \cite{pontifix_ref,Sule_2011}.
The Albe-Erhart-type BOP is based the original Brenner-formulation of BOP,
however, provides more flexibility in the potential
introducing further parameters.
The A-type BOP can be converted into Tersoff-formalism hence any code which
can handle the latter BOP can make use of A-BOP.
The advantage of A-BOP is the increased flexibility of the formalism
over the standard Tersoff-formalism.
Moreover, the interconversion of the Tersoff-formula into
the Brenner-type BOP (such as the A-BOP) is straightforward. 
\footnotetext[1]{The details of the interconversion can also be found
in an example potential file given in the recent releases of the code LAMMPS \cite{lammps} }

 We use in this paper the PONTIFIX code developed for fitting the 
Albe-type BOP \cite{pontifix_ref} for various binary compounds.
The obtained new parameter set has been tested by various molecular
dynamics codes (PARCAS \cite{parcas}, LAMMPS \cite{lammps}).
We find that although the new BOP for FeSi gives satisfactory
results, however, the stability of the B2 phase is notoriously
overestimated above the B20 phase which is the most stable form
of FeSi in nature under ambient conditions.
Therefore, we added a Coulombic term to the BOP in order to
provide a further flexibility in the potential and physically
to account for the partial ionic behaviour in this material
which could stabilize the B20 phase vs. the B2 one.
Similar extension of the BOP has been done recently for GaN \cite{GaN+C}.
The application of the derived new empirical potential
could prove valuable, since at best of our knowledge there is
no other empirical potential is available for FeSi.

%\nopage

%\definecolor{mycolor}{rgb}{0.9,0.9,0.0}
%\definecolor{mycolor}[cmyk]{0,0,1,0}

%\pagecolor{mycolor}

%\color{blue}{
%{\LARGE

%\underline{\bf The Albe-Erhart-type bond order potential:}\\ 
\section{The Generalized Albe-Erhart-type bond order potential}

%\bf{

 We give the generalized functional form of the Albe-Erhart-type bond order potential \cite{Albe,Erhart} which includes short and long range terms,
%\fbox{
\be
E=\sum_{ij,i>j} f_{ij}(r_{ij})[V_{ij}^R-b_{ij} V^A_{ij}(r_{ij})]+E_{long},
%\nonumber
%\label{radial}
\ee
%}
%}
where for the short range repulsive and attractive pair interactions
are the following, respectively,
%\fbox{\parbox{5cm}{\[
%\fbox{
\be
V_{ij}^R=\frac{D_0}{S-1} exp(-\beta \sqrt{2S}(r-r_0))
%\nonumber
\ee
%}

%\fbox{
\be
V_{ij}^A=\frac{S D_0}{S-1} exp(-\beta \sqrt{2/S}(r-r_0))
%\nonumber
\ee
%}

%\vspace{0.5cm}

%\fbox{
\be
b_{ij}=(1+\chi_{ij}^n)^{\frac{1}{2n}}
%\nonumber
\ee
%}

%\\
%\vspace{0.5cm}

%\fbox{
\be
\chi=\sum_{k(\neq i,j)} f_{ik}^c(r_{ik}) g_{ik}(\Theta_{ijk})  exp[2\mu_{ik}(r_{ij}-r_{ik})]
%\nonumber
\ee
%}

%\vspace{0.5cm}

%where the angular term $g(\Theta)$,
%\fbox{
\be 
g(\Theta)=\gamma \biggm(1+\frac{c^2}{d^2}-\frac{c^2}{d^2+(h+cos\Theta)^2}\biggm)
%\nonumber
\ee
%}
%\\
%\vspace{0.5cm}

\[ 
f_{ij}(r_{ij})= \left\{ \begin{array}{cc}
%{r@{\quad: \quad}1}
~~~~~~~~~ 1 & r \le R_c-D_c  \\ \frac{1}{2}-\frac{1}{2}sin[\frac{\pi}{2}(r-R_c)/D_c] & |r-R_c| \le D_c
 \\
0 & r \ge R_c+D_c 
\end{array} \right. \]
where $R_c$ is the short range cutoff distance.

 The long range part $E_{long}$ has been neglected in the
original formulation of the bond order formalism \cite{Albe,Erhart}.
\be
E_{long}=E_{Coul}+E_{Vdw}=\sum_{ij,i>j} \biggm[ \frac{1}{4 \pi \epsilon_0}\frac{q_i q_j}{r_{ij}} +\frac{C^{ij}_{Vdw}}{r_{ij}^6} \biggm]_{r_{ij} < r_{cut}}
\ee
where $q_i$, $q_j$, $R_c$, $r_{cut}$ and $C^{ij}_{Vdw}$ are net atomic charges, cutoff distances and van der Waals parameter for the BOP, Coulomb and van der Waals interactions, respectively.
The long range part allow us to extend the BOP for systems
with partial ionic character and also one can account for
long range effects excluded in the original BOP which cuts off
interactions at the first neighbors.

 We also study the effect of direct Coulomb interaction (BOP+C)
when the Van der Waals long range part $E_{Vdw}=0$,
\beqn
E^{BOP+C}=\sum_{ij,i>j} f_{ij}(r_{ij})[V_{ij}^R-b_{ij} V^A_{ij}(r_{ij})]_{r_{ij} < R_c} && \\ +E_{Coul}. 
\nonumber
%E^{BOP+C}=\sum_{ij,i>j} \biggm[ f_{ij}(r_{ij})[V_{ij}^R-b_{ij} V^A_{ij}(r_{ij})]_{r_{ij} < R_c}+ E_{Coul} && \\
% \frac{1}{4 \pi \epsilon_0}\frac{q_i q_j}{r_{ij}} \biggm|_{r_{ij} < r_{cut}}
%\biggm],
\label{bop+c}
\eeqn
%}
%} }
%where $q_i$, $q_j$, $R_c$ and $r_{cut}$ are net atomic charges and cutoff distances for the BOP and Coulomb interactions, respectively. 
The net point charges obtained from Bader's population analysis \cite{Bader,AIM} using
{\em ab initio} DFT calculations \cite{SIESTA}.
Instead of the diverging direct Coulomb sum in Eq. ~\ref{bop+c}
the damped shift forced coulomb (DSFC) sum method \cite{Fennel} has been used
which has been implemented by the author into the recent version of LAMMPS:
\beqn
\frac{1}{4 \pi \epsilon_0}\frac{q_i q_j}{r_{ij}} \approx
%V_coul = 1/rij + rij/r_cut^2 - 2/r_cut
\frac{q_i q_j}{4 \pi \epsilon_0} \biggm(\frac{1}{r_{ij}}+ && \frac{r_{ij}}{r_{cut}^2}
-\frac{2}{r_{cut}} \biggm),
\eeqn
where $r_{cut}$ is the applied cutoff distance for long range interactions.
The DSFC approach allows the treatment of periodic and non-periodic systems
while the originally implemented Ewald and pppm approaches
can be used for periodic systems only \cite{lammps}.
This extension of the original BOP allows us to account for the
partial ionic nature present in most of the binary compounds.
Although, it is rather difficult to obtain unambigously
the net charges $q_i$. The traditionally used Mulliken charges
are known to be exaggerated.
%are known to be exaggerated \cite{mulliken}.
It is widely accepted now that the net atomic charges obtained by
the Bader's decomposition scheme (atom in molecules, AIM) \cite{Bader}
are more relaible and is more or less free from spurious basis set dependence.
We calculate in the present paper the net charges using the SIESTA code \cite{SIESTA} and post calculations which generate Bader's charges.

 It could be usefull to note that 
 the obtained parameter set is compatible with the original Tersoff-formalism
built in e.g. widely available MD codes, such as LAMMPS \cite{lammps},
we give the formulas for transforming the parameters.
Note, that hereby we use the original symbols used by Tersoff \cite{Tersoff-Brenner}.
The radial part of the Tersoff potential is composed of the
following repulsive and attractive functions,
\be
V_{ij}^R=A \times exp(-\lambda_1 (r-r_0)),
\ee
\be
V_{ij}^A=B \times exp(-\lambda_2 (r-r_0)).
\ee
The required parameters $A, B, \lambda_1$ and $\lambda_2$ can be 
expressed using the Alber-Erhart parameters as follows:
\be
\lambda_1=\beta \sqrt{2S}, \lambda_2=\beta \sqrt{2/S}, \lambda_3=2 \mu=0.
\ee
\beqn
A=D_0/(S-1)*exp(\lambda_1*r_0), && \\
B=S*D_0/(S-1)*exp(\lambda_2*r_0), 
\eeqn
The parameters in the angular part are identical.
This conversion must be done when e.g. the LAMMPS code is used.
Example file can be found in the released packages of LAMMPS.
Moreover, the addition of the Coulomb part to the BOP core
is also straightforward using the "hybrid/overlay" option in the
code LAMMPS.

%%%%%%%%%%%%%%%%%%%%%%%%%%%%%%%%%%%%%%%%
% T1
%%%%%%%%%%%%%%%%%%%%%%%%%%%%%%%%%%%%%%%
%\LARGE{
\begin{table}[t]
\caption[]
{
The fitted parameters used in the Albe-Erhart type bond order interatomic potential for
the Fe-Si interaction
{\em The 2nd set of parameters} with the no use of Eqs (15)-(16) at the fitting prcodure.
Dimer properties were allowed to be reproduced badly.
No Pauling plot has been used for getting an initial guess.
}
%{\large
\begin{ruledtabular}
\begin{tabular}{ccc}
 \bf{Fe-Si} & BOP & BOP+C   \\
\hline
BOP-I & & \\
\hline
$D_0$ (eV)                   & 6.5588884 & 7.7457346 \\
$r_0$ ($\hbox{\AA}$)         & 1.5889110 & 1.5110803  \\
S                            & 1.9038691 & 1.6619156 \\
$\beta$ ($\hbox{\AA}^{-1}$)  & 1.0835049 & 1.2001554 \\
$\gamma$                     & 0.0809365 & 0.0798811 \\
c                          & 0.328786831 & 0.3327120 \\
d                          & 0.153064119 & 0.1570405 \\
h                         & -0.634457610 & -0.586481  \\
$R_c$ ($\hbox{\AA}$)         & 2.99671618 & 3.097295 \\ 
$D_c$ ($\hbox{\AA}$)         & 0.2   & 0.2     \\
n                          & 1.0 & 1.0 \\
$\mu$                      & 1.0 & 1.0   \\
\hline
BOP-II & & \\
\hline
$D_0$ (eV)                   & 3.0606066 & 3.1717432  \\
$r_0$ ($\hbox{\AA}$)         & 2.0493522 & 2.0174542  \\
S                            & 3.9625636 & 3.7495785 \\
$\beta$ ($\hbox{\AA}^{-1}$)  & 0.7292032 & 0.9780126  \\
$\gamma$                     & 0.0780215 & 0.0716019\\
c                            & 0.3142206 & 0.3097108 \\
d                            & 0.1546207 & 0.1560145 \\
h                            & -0.6230591 & -0.8560120 \\
$R_c$ ($\hbox{\AA}$)         & 3.0169527 & 3.1056216 \\
%$D_c$ ($\hbox{\AA}$)         & 0.2 &   0.2 \\
%n                          & 1.0 & 1.0 \\
%$\mu$                      & 1.0 & 1.0   \\
\hline
BOP-IIb & & \\
\hline
%\multicolumn{3}{l}{
%fit to heat of mixing and to $E_c$ (lammps_bop-vi)
%} \\
\hline
$D_0$ (eV)                   & 2.4282727 & 2.33410201 \\
$r_0$ ($\hbox{\AA}$)         & 2.0330640 & 2.07023841  \\
S                            & 3.8251348 & 4.13277821 \\
$\beta$ ($\hbox{\AA}^{-1}$)  & 0.8244547 & 0.89583103 \\
$\gamma$                     & 0.068166344 & 0.069052916 \\
c                            & 0.300448478  & 0.30345365 \\
d                            & 0.159001935 & 0.157617464 \\
h                            & -0.92396567 & -1.019548197 \\
$R_c$ ($\hbox{\AA}$)         & 3.170028434 & 3.143001937 \\
\hline
BOP-III & & \\
\hline
$D_0$ (eV)                   & 6.4544830 & 5.2257579  \\
$r_0$ ($\hbox{\AA}$)         & 1.4586364 & 1.6094938  \\
S                            & 1.6830565 & 1.5510515  \\
$\beta$ ($\hbox{\AA}^{-1}$)  & 1.1313769 & 1.2645078  \\
$\gamma$                     & 0.076970476 & 0.066245953 \\
c                            & 0.33214478 & 0.30394206 \\
d                            & 0.15342928 & 0.16913177 \\
h                            & -0.6523133 & -0.67888934 \\
$R_c$ ($\hbox{\AA}$)         & 3.02677686 & 3.00072535 \\
\hline
%---------------------------------------------------------------
\end{tabular}
\end{ruledtabular}
%}
%\footnotetext[1]{
%}
\label{T1}
\end{table}

%%%%%%%%%%%%%%%%%%%%%%%%
% T2
%%%%%%%%%%%%%%%%%%%%%%%%
%\begin{table*}[tb]
\begin{table}[t]
\raggedright
%\sidecaption
\caption[]
{
The summary of the
basic results obtained for the dimer FeSi.
Experimental data is from ref. \cite{fesi_exp},
theoretical data is taken from ref. \cite{fesi_theo}. 
}
\begin{ruledtabular}
\begin{tabular}{cccc}
        &    Expt. & Theory & BOP  \\
\hline
 BE         &  3.037 $\pm$ 0.259 & 2.236, 2.743$^a$, 3.217$^b$ &    \\
 r$_0$      & & 2.19-2.23$^c$, 2.011-2.091$^a$ &   \\
 $\omega_0$ & & 315.6$^d$, 248.9$^d$ &   \\
\end{tabular}
%\end{ruledtabular}
%\footnotetext[1]{
\footnotemark*{
present work: DFT calculation using the molecular G03 code \cite{G03} and
PBE exchange-correlation funcional or hybrid functionals such
as B3LYP.
$^b$ Obtained by the Quantum Espresso suit plane-wave DFT code (PW) \cite{QE}
using the HSE functional with Hartree-Fock exchange.
BE: bonding energy (eV), $r_0$: equilibrium distance ($\hbox{\AA}$),
$\omega_0$: ground state oscillation frequency (cm$^{-1}$).
%No experimental datqa is found for $r_0$ and $\omega_0$.
%However, we give estimated values after taken the data of
%similar molecules (e.g. NiSi). Most of the metal silicides
%possess data in this regime.
$^b$ ref. \cite{fesi_theo},
$^c$ present work: QCISD(T)/LanL2DZ, B3LYP/LanL2DZ, using G03.
}
\label{T2}
\end{ruledtabular}
%\end{table*}
\end{table}

%\LARGE{
\begin{table}[t]
\label{T4}
\caption[]
{
The fitted parameters used in the bond order interatomic potential for
the Fe-Si interaction transformed into the Tersoff's formula.
The radial part is different only. The angular parameters are the
same have shown in Table I.
}
%{\large
\begin{ruledtabular}
\begin{tabular}{ccc}
\hline
 \bf{Fe-Si} & Tersoff & Tersoff+C   \\
\hline
 BOP-I &  &   \\
\hline
$A$ (eV)                        & 208.785964 & 319.282671   \\
$B$ (eV)                        & 80.664727 & 142.193959 \\
$\lambda_1$ ($\hbox{\AA}^{-1}$) & 2.114289 & 2.18804875 \\
$\lambda_2$ ($\hbox{\AA}^{-1}$) & 1.110522 & 1.31658233 \\
\hline
 BOP-II &  &   \\
\hline
$A$ (eV)                        & 69.3743499 & 256.229868    \\
$B$ (eV)                        & 11.8357598 &  18.2744415 \\
$\lambda_1$ ($\hbox{\AA}^{-1}$) & 2.0528238 & 2.67824732  \\
$\lambda_2$ ($\hbox{\AA}^{-1}$) & 0.51805447 & 0.714279566  \\
% 1.31658233  142.193959  3.09729471  .2  2.18804875  319.282671
\hline
 BOP-IIb & lammps-bop-vi &   \\
\hline
$A$ (eV)                        & 88.6557396  & 154.101787     \\
$B$ (eV)                        & 11.0478628 &  11.1876165 \\
$\lambda_1$ ($\hbox{\AA}^{-1}$) & 2.28036929  & 2.57550353 \\
$\lambda_2$ ($\hbox{\AA}^{-1}$) & 0.596153974 & 0.623189388  \\
%  .623189388  11.1876165  3.14300194  .2  2.57550353  154.101787
%  1.23331197  96.1123514  3.02677686  .2  2.07573374  195.135348
\hline
 BOP-III & lammps-bop-v &   \\
\hline
$A$ (eV)                        & 195.135348  & 341.761224    \\
$B$ (eV)                        & 96.1123514  & 148.343114  \\
$\lambda_1$ ($\hbox{\AA}^{-1}$) & 2.07573374  & 2.22715079  \\
$\lambda_2$ ($\hbox{\AA}^{-1}$) & 1.23331197  & 1.43589738  \\
%  1.43589738  148.343114  3.00072535  .2  2.22715079  341.761224
\hline
%---------------------------------------------------------------
\end{tabular}
\end{ruledtabular}
%}
\footnotetext[1]{
This Table gives the parameters in those form which can be used
in tersoff potential files used by e.g. LAMMPS.
}
\end{table}

\section{The parameter fitting procedure}

\subsection{The preparation of the initial guess}

 The selection of the initial guess for the parametrization procedure
is carried out as follows:
The ground state oscillation frequency of the dimer,
\be
\beta=\frac{1}{2} \omega_0 \sqrt{2\mu/D_0},
\ee
where $\omega_0$, $\mu$ and $D_0$ are the
zero point vibration frequency, reduced mass
and dissociation energy of the dimer, respectively.
The values shown in Table III. have been used to calculate
the initial guess $\beta$ for the radial part of the BOP.
The adjustment of parameter $S$ has been done by tuning
the Pauling relation for bond order (following the method
proposed in recent publications \cite{ZnO,pontifix_ref}),
\be
E_b= D_0 exp[-\beta \sqrt{2S} (r_b-r_0)],
\ee
where $r_b$ is the first neighbor distance in
various polymorphs, and $r_0$ is the dimer interatomic distance.
%$r_b$ can be given as the weighted average of the
%next nearest neighbor distances within cutoff distance obtained from
%the rdf file: $r_b = \frac{1}{W} \sum_i d_i w_i$,
%where $d_i$, $w_i$ and $W=\sum_i w_i$ are the interatomic
%distances, the corresponding weight in the histogram and
%the sum of the weights
%as appearing in the rdf histogram file produced e.g. by the
%code LAMMPS.
%Instead we adjusted it manually, and selecting a starting $S \approx 1.8$
%we get a reasonable parameter set.

%%%%%%%%%%%%%%%%%%%%%%%%%%%%%%
% F1
%%%%%%%%%%%%%%%%%%%%%%%%%%%%%
\begin{figure}[hbtp]
\begin{center}
\includegraphics*[height=6cm,width=8cm,angle=0.]{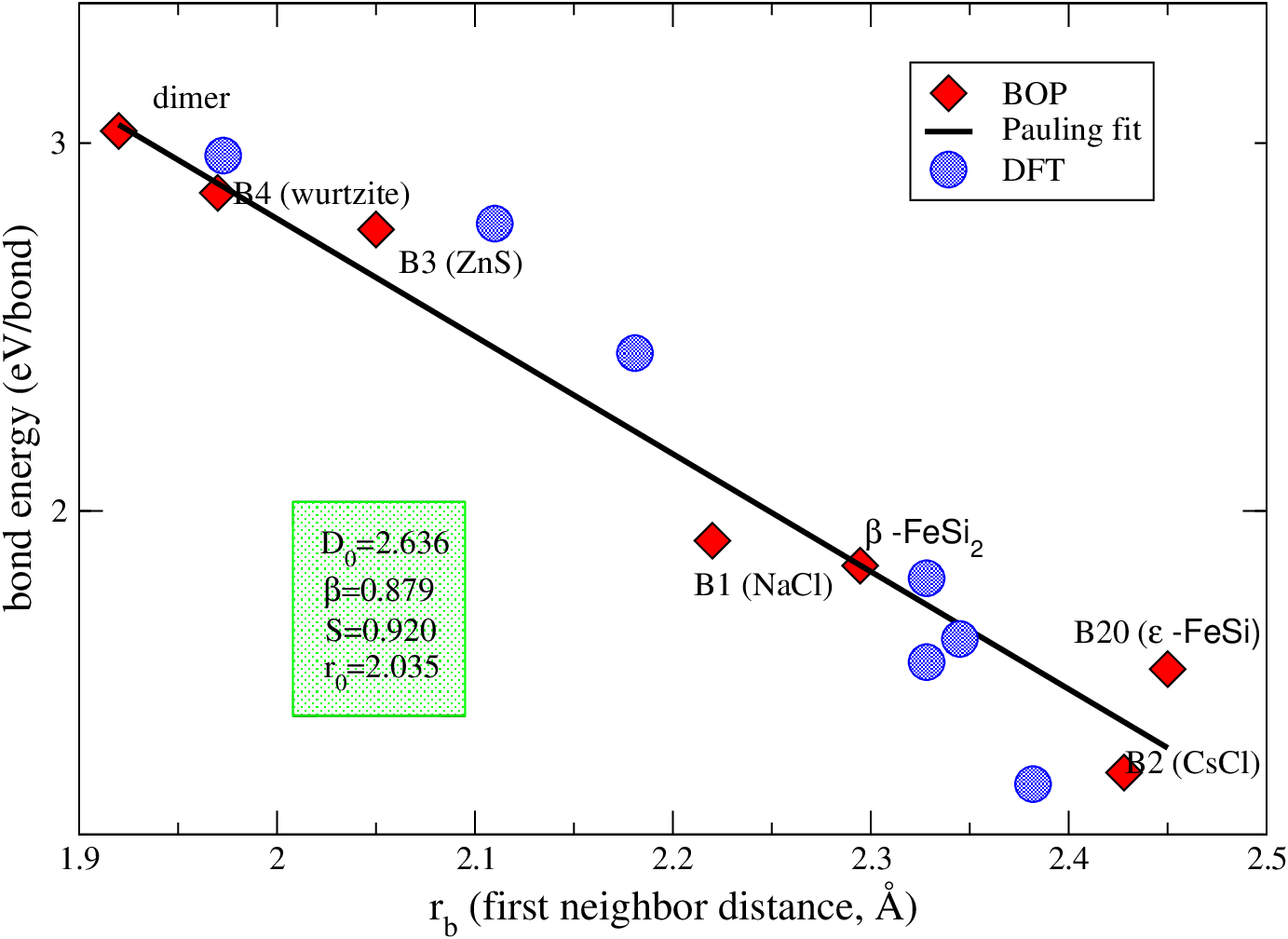}
\caption[]{
Pauling plot (Eq. (16)) for BOP-II comparing data obtained from the
analytic bond order potential (first parameter set) for various polymorphs.
Few DFT results are also shown obtained by the author.
The obtained fitted values of the Pauling bond order
expression are also given. These values were used
as initial guess for the radial part during the
parametrization of the BOP.
%The fitted straight line helps to demonstrate the
%scatter of the data.
}
%\label{compare}
\end{center}
\footnotemark{
The Pauling bond order expression has been fitted
to the DFT points shown. 
The obtained $D_0$, $\beta$ and $S$ parameters were used
as initial guess for the parametrization of the BOP.
$r_0$ has been kept fixed at $r_0=2.09$ $\hbox{\AA}$
which has been found by DFT calculations in the present
work (see Table III.).
}

\label{F1}
\end{figure}
%------------------------------------------------------
The obtained Pauling plot (semi-logarithmic plot)
and the initial guess for the radial parameters are shown in Fig. 1.
As can be seen, the fit to the DFT data results in a somewhat
different curve than the one obtained by the final parametrized BOP.
The reason could be that the DFT data is not fully consistent
with the BOP. Unfortunatelly, experimental data is not available
for the cohesive properties of the various polymorphs, hence, it is hard
to consider the accuracy of the obtained DFT results.
The fit of the Pauling relation given by Eq. (16) results in
the fit values shown in Fig. 1.
These values were used as an initial guess for the parametrization procedure.

 In fact, we obtained two sets of parameters, BOP-I and BOP-II.
The overall performance of the two force fields, as it has been shown in Table (4) is rather similar
except for the melting point which used to be critical
for the BOP \cite{Albe}.
BOP-II has been generated in a standard way hence we include
this potential just for comparison and for the analysis of the
results.
We started from the Pauling curve in this case (BOP-II).
However, we also derived another set of parameters for BOP.
In this case (BOP-I) we did not stick the initial parameters to the
dimer and we let the BOP to be fitted to the training set only
which, however, also inlcudes the dimer.

  The basic difference between the two parameter fitting procedure is,
however, in the following.
In the case of BOP-II one can simply generate the parameters
%%%%%%%%%%%%%%%%%%%%%%%%%%%%%%
% F2
%%%%%%%%%%%%%%%%%%%%%%%%%%%%%
\begin{figure}[hbtp]
\begin{center}
\includegraphics*[height=6cm,width=8cm,angle=0.]{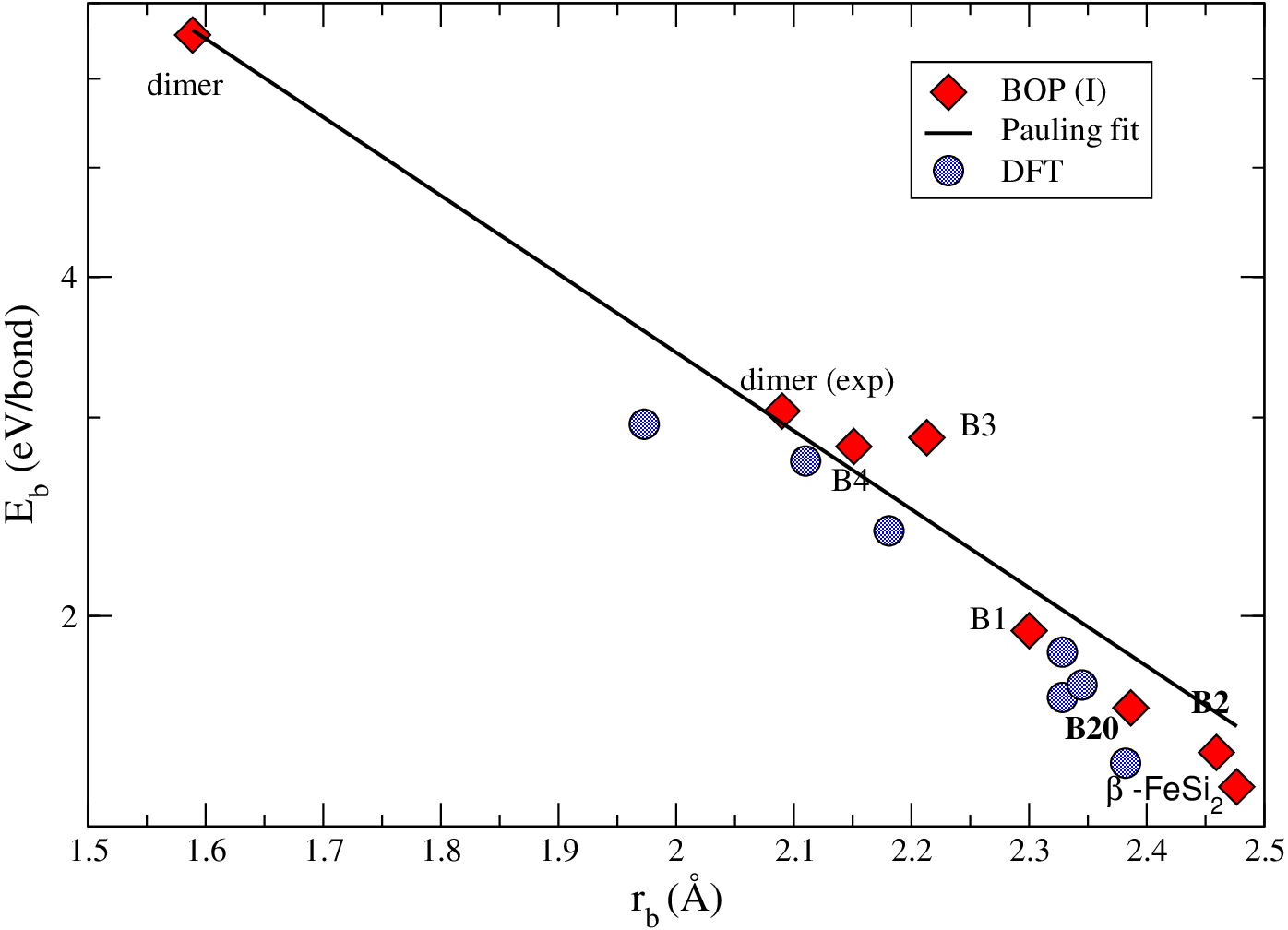}
\caption[]{
The
Pauling plot (Eq. (16)) for FeSi for the BOP-I parameter set comparing data obtained from the
analytic bond order potential (first parameter set) for various polymorphs.
Few DFT results are also shown obtained by the author.
The obtained Pauling fit parameters were not used for BOP-I
as initial guess and is shown here just for guiding the eye.
Instead, the initial guess parameters have been obtained on a trial-and-error
basis and set in manually.
This way of parametrization is, however, leads to a somewhat
tedious procedure with repeated samplings of the configuration space
of the parameters.
Typically few tens of a trials with setting in reasonable values
for the radial parameters and with the recycling
of the output parameters (when we find better and better values).
In the case of convergence one can refine in this way the
obtained parameter set.

%The obtained fitted values of the Pauling bond order
%expression are also given. These values were used
%as initial guess for the radial part during the
%parametrization of the BOP.
%The fitted straight line helps to demonstrate the
%scatter of the data.
}
%\label{compare}
\end{center}
\footnotemark{
}
\label{F2}

\end{figure}
%------------------------------------------------------
with a single iteration run with the PONTIFIX code.
For BOP-I, a series of iterations have been employed recursively
until a satisfactory result obtained.
It turned out that this procedure moves the parameters towards
larger $D_0$ and shorter $r_0$ than the equilibrium dimer values.
Typically we use few tens of reiterations of intermediate
parameter sets until convergence reached in the "super-iterative"
fitting procedure.
At the end of each iterative steps (Levenberg-Marquardt least-squares algorithm \cite{Press}) the obtained temporal parameters have been fed back to
the algorythm until final convergence is reached.
Hence, using a series of iterative steps instead of a single one,
one can scan the parameter space for a global minimum.
%Simply, we do not set in in this case parameter $\beta$ by Eq. (15)
%and we do not use Eq. (16) for finding parameter $S$, as it has
%been proposed in the original BOP recipe \cite{pontifix,PONTIFIX}.
%Using this second set we get a BOP which reproduces surprisingly well the
%dimer properties, and also seems to perform better than 
%BOP-II parameter set.
While BOP-II gives too large melting point ($T_m \approx 2300 \pm 100$ K), the second set
gives a rather satisfactory one for the B20 phase ($T_m \approx 1550 \pm 100$ K).
The overestimated melting point is a well known problem of
BOP from previous publications (see. e.g. ref. \cite{Albe}).
This problem was mostly known for semiconductors \cite{Albe}.
Since this is an important issue, it could be useful
to figure out while BOP-I parameter set gives much lower $T_m$ than
BOP-II.
Our guess is that the initial Pauling-constraints on the initial parameter set
(initial guess) put by the Pauling relation
will scan those part of the complicated parameter space which prefers
notorouisly the overbinding of various phases.
%%%%%%%%%%%%%%%%%%%%%%%%%%%%%%
% F3
%%%%%%%%%%%%%%%%%%%%%%%%%%%%%
\begin{figure}[hbtp]
\begin{center}
\includegraphics*[height=6cm,width=8cm,angle=0.]{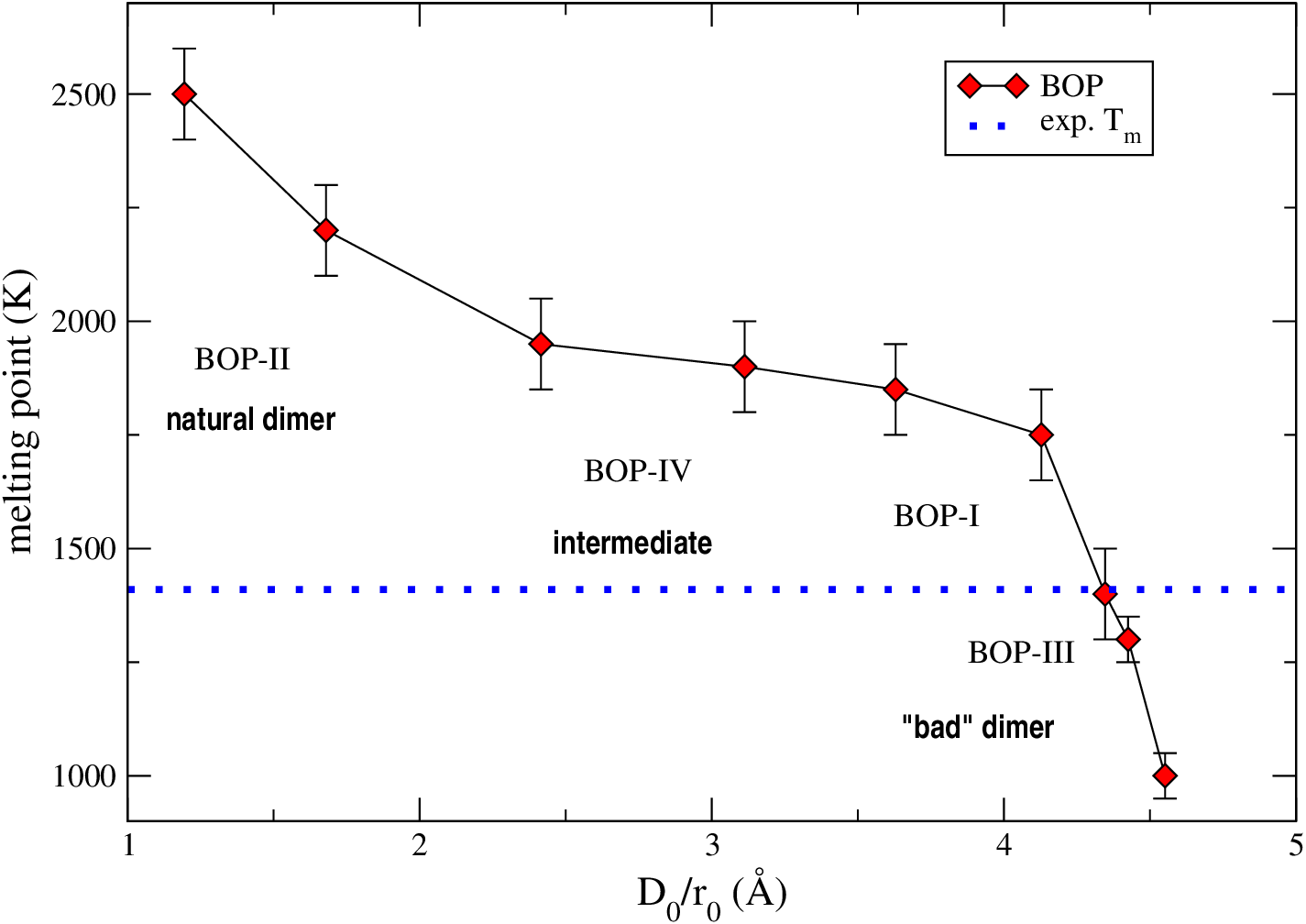}
\caption[]{
The melting point against the dimer ratio $D_0/r_0$.
}
%\label{compare}
\end{center}
\footnotemark{
}
\label{F3}
\end{figure}
%%%%%%%%%%%%%%%%%%%%%%%%%%%%%%

When this constraint is released, we allow the exploration of
further "hidden" parts of the parameter space.
In particular, if we allow shorter dimer equilibrium distance $r_0$, the
BOP will provide melting at around the experimental
temperature. 
We find that the proper melting behavior occurs when $D_0$ is chosen to be much higher than
the experimental one together with a much shorter $r_0$.
Hence, the selection of unphysical $D_0$ and $r_0$ leads to a
more relaible BOP which performs also well for the dimer (see Fig. 2).
In particular, BOP-I gives bonding energy ($E_b$) around
the physical (experimental, $E_b \approx 3$ eV/bond) one together with a reasonable
$r_b \approx 2.1$ $\hbox{\AA}$ (see the solid line of the Pauling fit). 
%Hence loosing accuracy in the one side (dimer properties) we gain
%accuracy on the other side in the solid phase with higher coordination numbers.

\subsection{The parametrization and the training set}

 The BOP potential has been parametrized using an extended training set of
various structures (dimer, B1 (NaCl), B2 (CsCl), B3 (ZnS), B20 (eps-FeSi), $Fe_3Si$ (L12a), $FeSi_3$ (L12b) phases for FeSi).
The parametrization procedure has been carried out using the PONTIFIX code
developed by P. Erhart and K. Albe \cite{PONTIFIX,pontifix_ref}.
%The parameters of FeSi have simulataneously been fitted together
%with the elemental Si and Fe parameters.
For the elements similar data base has been used given by
Albe and Erhart \cite{pontifix_ref}.
{\em ab initio} SIESTA \cite{SIESTA} calculations were used to determine the cohesive energies
of various structures.
   A Levenberg-Marquardt least-squares algorithm \cite{Press} has been implemented
in Pontifix to find a combination of parameters which minimizes the deviation
between the properties in the fitting database and the properties predicted by the potential.
Parameter sets for different interaction types can be fitted simultaneously.
The fitting database encompassed the bond lengths and energies of various structures as well as elastic
constants.
Subsequent fitting trials and refinements are exploited (time to time on a trial and error basis) until a satisfactory
parameter set has been obtained for the most stable B20 phase.
The long-range terms have also been implemented in the modified
version of the PONTIFIX code which allows
the explicit parametrization of the generalized BOP.
Results will be shown for the BOP and BOP+C parameter sets.
The parameter sets have been tested for the various polymorhps of FeSi
(B1, B2, B3, B20, $\beta-FeSi_2$, $\gamma-FeSi_2$)
using the modified version of the PARCAS code \cite{parcas}
and the LAMMPS code \cite{lammps}.
The DSCF method has also been implemented in the LAMMPS 
as well as in the PONTIFIX codes. 
The main advantage of the DSCF approach beyond its simplicity is that it can also be used
for non-periodic systems.
%The BOP formalism is available also in the code LAMMPS \cite{lammps}.

\subsection{Multiple set of parameters for various application fields}

 We realized 
that instead of developing only a universal potential with a relatively weak
accuracy it is better to develop a few sets of parameters which optimized
to different properties of the training set.
In particular, it is hard to achieve the dimer properties
accurately together with the solid state.
It turned out that keeping the radial part accurately fitted to the dimer
(as it has been proposed for the parametrization of the Alber-Erhart-type
BOP) leads to the serious overestimation of the melting point
while other properties remain acceptable.
Since in the case of FeSi it is important to 
simulate accurately $T_m$ due to special application fields related to
this quantity (high pressure phases of FeSi and melting see e.g. ref. \cite{Vocadlo}).
Keeping this in mind 
we release in this paper few sets which were parametrized independently:
\\
BOP-I: The cohesive energies are fitted to the {\em ab initio} training set with short $r_0$ and
large $D_0$ ($r_0 \ll 2.0$ $\hbox{\AA}$, $D_0 \gg 3.0$ eV)
using scaled down {\em ab initio} DFT cohesive energies. 
Poor dimer property, good melting point and lattice constant, 
acceptable elastic properties, proper phase order of the B20 and B2 polymorphs
for the BOP+C variant.
The improvement of $T_m$ is also significant for the BOP+C variant over
the BOP.
\\
BOP-II: Fitted to the ab initio training set (BOP-IIa) or to experimental heat of formation (BOP-IIb) with natural $r_0$ and
$D_0$. ($r_0 \approx 2.1 \pm 0.1$ $\hbox{\AA}$, $D_0 \approx 3.0 \pm 0.5$ eV).
This potential provides too large melting point of $T_m > 2200$ K instead
of the experimental $T_{m,exp} \approx 1450 \pm 50$ K.
$T_m$ remains high for
BOP-IIb, although the cohesive energies are lowered significantly as for BOP-III. 
It can be taken for granted that good dimer properties keeps 
$T_m$ too high.
There seems to be no solution for this paradoxon.
The overall performance is the following: Good at the dimer, overestimated cohesive energies (BOP-IIa),
acceptable elastic properties.
Bad phase order in sign for B20 vs. B1 polymorphs ($\Delta H_{B20-B2} \approx 0.15$ eV/atom instead of $\Delta H_{B20-B2,exp} \approx -0.25$ eV/atom) \cite{enthalpy}.
We do not reccomend this parameter set for applications and we include it
just for demonstrative purpose.
\\
BOP-III: Fitted to experimental heat of formation ($\Delta H$) with short $r_0$ and
large $D_0$ ($\Delta H \approx 0.5 \pm 0.2$ eV/atom) in such a way
that the ab initio cohesive energies are scaled down by $\sim 0.95$ eV/atom
in order to reproduce experimental $\Delta H$ \cite{thermo_tmsil}.
BOP-III is not recommended for the dimer (bad dimer regime). Melting occurs at somewhat low temperature of
$T_m \approx 1250 \pm 50$ K, although it is not far from $T_{m,exp} \approx 1410$ K.
Cohesive energies and $\Delta H$ are realistic as being fitted to them.
Phase order for B20 vs. B1 polymorphs is also in the right side 
($\Delta H_{B20-B2} \approx -0.05$ eV/atom).
%%%%%%%%%%%%%%%%%%%%%%%%%%%%%%
% F4
%%%%%%%%%%%%%%%%%%%%%%%%%%%%%
\begin{figure}[hbtp]
\begin{center}
\includegraphics*[height=6cm,width=8cm,angle=0.]{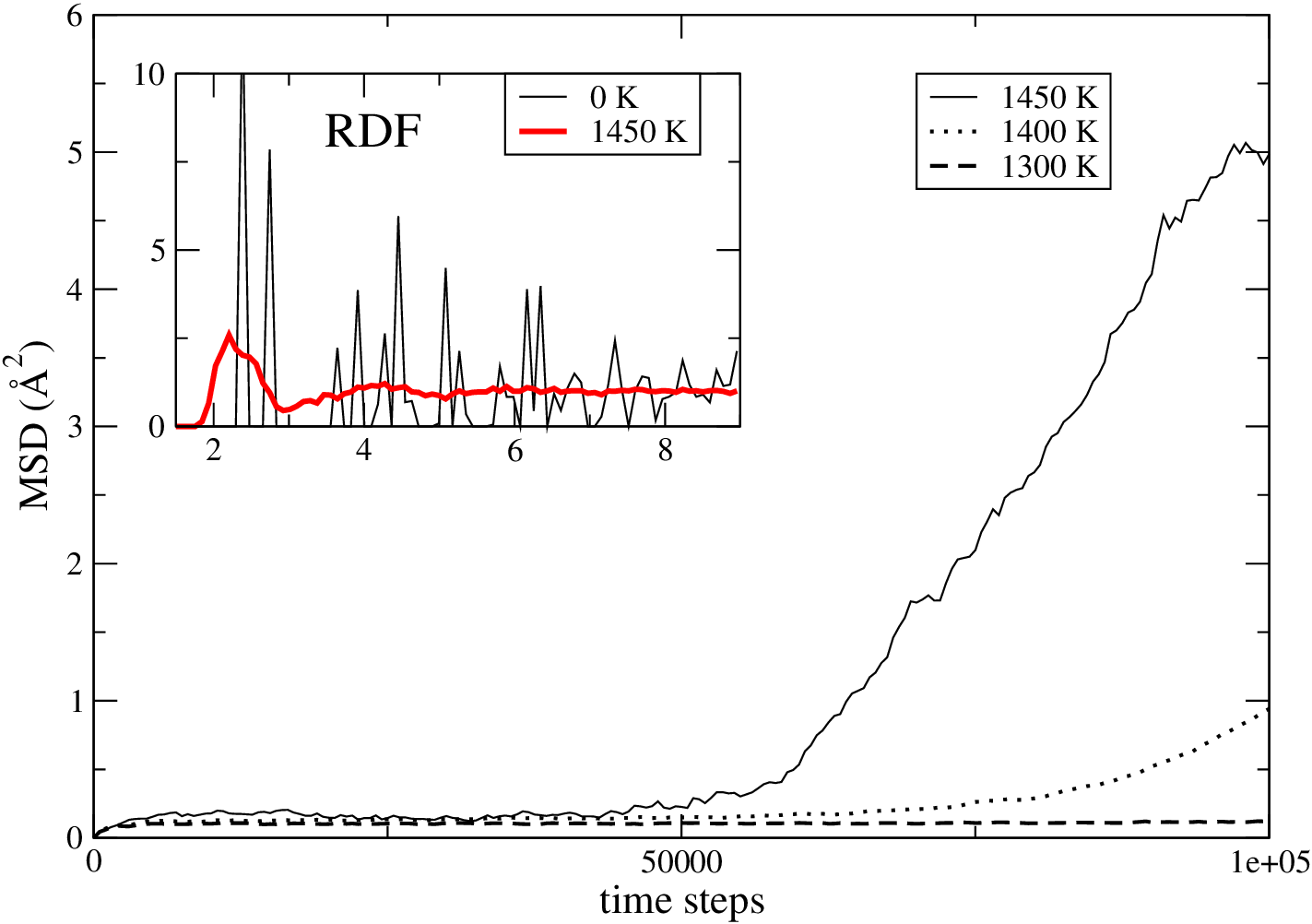}
\caption[]{
The mean square of displacements (MSD) vs. time steps
for the B20 system using the BOP-III+C parameter set
at various temperatures.
{\em Inset}: The radial density function (RDF) is also
shown for 0 K and at around the melting point. 
}
\label{compare}
\end{center}
\footnotemark{
}
\label{F4}
\end{figure}
%%%%%%%%%%%%%%%%%%%%%%%%%%%%%%%%%%%%%%%%%%%%%%%%%%%%%%%
The improvement of $\Delta H_{B20-B2}$ is also remarkable for the
BOP+C variant ($\Delta H_{B20-B2} \approx -0.13$ eV/atom).
As an overall conclusion we find that BOP-III performs
rather well and is suitable for various application fields related to FeSi.
\\
BOP-IV: Search for middle range parameters
($1.7 < r_0 < 2.0$ $\hbox{\AA}$, $D_0 > 3.0$ eV).
The combination of properties produced by BOP-I, BOP-II and BOP-III.
This leads to intermediate values for $r_0$ and $D_0$.
$\Delta H$ also becomes larger than the upper limit of the
measured values.
Unfortunatelly, it has been found that in this parameter regime
$T_m$ remains still too high at around $T_m \approx 2000 \pm 100$ K.
This parameter regime proved to be unsuccessfull, hence, we
do not show results in the rest of the paper.
\\

 Morever, each set is released with BOP and BOP+C variants.
Theferfore, finally we end up with 8 parameter sets.
One might think it is more than reasonable, however,
we argue that this "zoo" of force fields make it possible
to select the most appropriate one for the problem to be studied.
One has to keep in mind that empirical force fields are not unique
and the parameter space contains unexplored optimal regions which
could be similar to each other in performance.

\subsection{ab initio DFT results for the polymorphs}

 Since the availability of experimental results for the cohesive properties
of FeSi is rather limited we calculated the cohesive energy (binding energy per
atom) of various polymorphs using the SIESTA code \cite{SIESTA}.
The unit cells were optimized by the variable cell approach using
standard DZP basis for Fe and an optimized TZP quality basis for
Si together with standard pseudopotentials available from the homepage
of SIESTA. $3 \times 3$ Monkhorst k-grid has been used.
The obtained results are summarized in Table IV.

 One has to keep in mind, however, that the {\em ab initio} DFT
cohesive energies might be exagerated, which is due to the
notorious overbinding behavior of LDA and GGA exchange-correlation
functionals \cite{overbind}.
Therefore, we correct the obtained $E_c^{DFT}$ values by
$\Delta E_c \approx \Delta H_{exp}$.
One can estimate the "experimental" $E_c$ of the B20 phase using
the measured formation energy of $\epsilon$-FeSi, which is
in the range of $\mid \Delta H_{exp} \mid \approx 0.2-0.65$ eV/atom \cite{thermo_tmsil}:
$E_{c,est} \approx (\Delta H_{exp}+E_{c,Fe}+E_{c,Si})/2 \approx 4.995$ eV/atom,
where $E_{c,Fe}=-4.28$ eV/atom and $E_{c,Si}=-4.63$ eV/atom are the measured cohesive energies of the
constituents, respectively.
We find than that the DFT GGA model overestimates the cohesive energy by some
$\Delta E_c \approx \mid E_c^{DFT}-E_{c,est} \mid \approx 5.928-4.995=0.933$ eV/atom.
$\Delta E_c$ is used than to correct the {\em ab initio} DFT cohesive energies
for fitting purpose.
Hence in the fitting data base we decreased the $E_c^{DFT}$ values by $\Delta E_c$.

%%%%%%%%%%%%%%%%%%%%%
% T4
%%%%%%%%%%%%%%%%%%%%%

%\begin{longtable}
%\begin{table*}[tb]
\begin{table*}
\raggedright
%\sidecaption
%\begin{table}[t]
%\label{T4}
\caption[]
{
ab initio results for various polymorphs of FeSi
(spinpolarized PBE) obtained for cohesive energy $E_{cohes}$ (eV/atom),
equilibrium lattice constant (${a_{latt}}$)
and net atomic charges/Si atom (Mulliken charges $q_{Si,M}$ and
Bader's charges using the Voronoi analysis, $q_{Si,B}$).
}
\begin{ruledtabular}
%\begin{minipage}{0cm}
\begin{tabular}{cccccc}
%\hline \hline
 polymorph & symbol & $E_{cohes}$ (eV)\footnote[1] & ${a_{latt}}$ ($\hbox{\AA}$) & $q_{Si,M}$ & $q_{Si,B}$  \\
\hline
 dimer \footnote[2] &                & -2.668  & 2.106 & -0.24 & -0.180  \\ 

 dimer \footnotemark[3] &          & -2.745  & 2.11 & -0.273 & \\ 
 dimer \footnotemark[4] &          &   & 2.00 & -0.24  &   \\ 
 B20 & $\epsilon$-FeSi  & -5.928 (-8.370) & 4.467 & -0.563 & -0.15 \\
 B2 & CsCl             & -5.920 (-7.415) & 2.751 & -0.338 & -0.231 \\
 B1 & NaCl            & -5.181 & 4.950 & -0.705 & -0.134 \\
 B3 & ZnS ?             & -5.918 & 2.758 & -0.344 & \\ 
 Fe$_{3Si}$, & L12a        & -5.991 & 3.595 & -0.923 & -0.359 \\
 Fe$_{3Si}$  & DO3, AlFe$_3$        & -3.580 & 4.412 & -0.153 &  \\
 FeSi$_3$    & L12b,CuAu$_3$ & -4.745 & 3.635 & -0.120 & -0.09 \\
 $\gamma$-FeSi$_2$ & C1, CaF$_2$  & -3.062 & 5.3 & -0.295 & \\
% $\alpha$-FeSi$_2$ (tP3)  & & & & \\
 $\beta$-FeSi$_2$ & oC48  & -5.571 & 9.816, 7.745, 7.813 & & \\
%\hline
%---------------------------------------------------------------
\end{tabular}
%\end{minipage}
%\end{ruledtabular}
%\footnoterule
%\footnotetext[1]{
%\footnotemark{
%\multicolumn{5}{l}{
\footnotetext*{
The cohesive energy of $Fe_nSi_m$ is $E_{cohes}=\frac{1}{n+m} \biggm( E_{tot}^{n,m}-n E_{Fe}-m E_{Si} \biggm)$, 
where $E_{tot}^{n,m}$ is the calculated 
total energy of the $Fe_nSi_m$ system and $E_{Fe}$ and $E_{Si}$ are the corresponding single atomic energies.
} 
\\
\footnotetext*{
Obtained by the fully periodic code SIESTA \cite{SIESTA}.
The results obtained by the fully periodic SIESTA code using spinpolarized
and spin-unpolarized DFT calculations with the PBE xc-functional
using $3 \times 3$ Monkhorst k-grid and TZTPF long basis, optimized for bulk
Si \cite{basis_Si} and a standard built-in DZP basis set for Fe.
We also give cohesive energies for the B2 and B20 polymorphs in parentheses
obtained by the fully periodic plane-wave QUANTUM ESPRESSO code \cite{QE} using
the HSE functional with Hartree-Fock exchange ($3 \times 3$ Monkhorst k-grid,
55 Ry plane wave cutoff, spin restricted).
}
\\
\footnotetext*{
Obtained using the molecular code Gaussian (G03) \cite{G03}
using PBE/6-311G(d) xc-functional and basis set.
}
\\
\footnotetext*{
Also with G03 using UB3LYP/6-311G(d).
}
\\
\footnotetext*{
\cite{Hafner},
Net atomic charges obtained by the Mulliken analysis (SIESTA, G03) and by
natural population analysis built in G03.
}
\end{ruledtabular}
\end{table*}
%\end{longtable}

%%%%%%%%%%%%%%%%%%%%%%%%
% T5
%%%%%%%%%%%%%%%%%%%%%%%%
%\begin{table*}[tb]
\begin{table*}
\raggedright
%\begin{table}[t]
%\sidecaption
%  \newcommand{\spr}[1]{\multicolumn{1}{c}{#1}}
% \centering
\caption[]
{
The comparison of BOP-I and BOP-II force fields
with each other and with density functional (DFT) and
experimental results for the B20 ($\epsilon$-FeSi) and B2 (CsCl)
polymorphs.
In both cases the short-range only (BOP) and the
short+long range BOPs (BOP+C) are compared.
BOP-I corresponds to the parameter sets obtained for
the "bad-dimer" case. BOP-II has been determined by
using the standard Pauling process for fitting the radial part.
}
\begin{ruledtabular}
%  \label{T5}
%  \begin{tabular}{ll*{9}l}
\begin{tabular}{cccccccccc}
%\hline \hline
%\toprule
  & dim & BOP-I & BOP-I+C & BOP-II & BOP-II+C & DFT & EXP \\
\hline
  & & & & & & & \\
 B20 ($\epsilon$-FeSi) & & & & & & &   \\
  && & & & & & \\
%-4.9329908    4.46532
%-4.8921645    4.46782
 $a_{latt}$ & $\hbox{\AA}$  & 4.465  & 4.468 & 4.490 & 4.491 & 4.75 & 4.489\footnotemark[1]  \\
 $V$ & $\hbox{\AA}^3$  &   &   &  &  & & & 90.19
%\footnote[1]{ denotes data which is not available from the literature or
%we could not derive by DFT calculations (e.g. $T_m$),
%DFT results obtained in the present work using the SIESTA package \cite{SIESTA}.
%Details of the calculations are given in Table ~\ref{T4}.
%}
\\
 $E_{cohes}$ & eV/atom     & -4.933 & -4.892 & -4.841 & -4.747 & -5.928 & -4.995 (est) \\
$|\Delta H|$ & eV/atom      & 0.239 & 0.219 & & & 0.687 & 0.28-0.64$^b$   \\
 $\gamma$ & eV/atom      & 0.54 & 0.53 & 0.54 & 0.54 & n.a. & n.a. &    \\
 $T_{melt}$ & K            & $1650 \pm 50$ & $1600 \pm 50$ & $2800 \pm 100$ & $> 2000$ & n/a & 1410$^b$, 1473$^c$  \\ 
  $B$  & GPa                & 193.4 & 232.7 & 224.9 & 232.7 & 180 & 160-209$^c$ \\
%  $B^{'}$  &               &  &  &  &  & & 4.5 \\
$c_{11}$ & GPa            & 281.1 & 346.5 & 296.8  & 316.1 & 488$^d$ & 316.7$^e$, 346.3$^f$ \\
$c_{12}$ & GPa              & 149.6 & 175.8 & 178.1 & 182.8 & 213 & 112.9, 139.2 \\
$c_{44}$ & GPa              & 90.2 & 82.9  & 106.7 & 112.2 & 125 & 125.2, 105.8  \\

  & & & & & & & \\
\hline \hline
%%%%%%%%%%%%%%%%%%%%%%%%%%%%%
% kj/mol to eV : 0.01036410
%%%%%%%%%%%%%%%%%%%%%%%%%%%%%
%\bottomrule
  & & & & & & & \\
 B2 (CsCl) & & & & & & & \\
  & & & & & & & \\
 & $a_{latt}$ ($\hbox{\AA}$)  & 2.801 & 2.803  & 2.798 & 2.801 & 2.751 & 2.83   \\
 $E_{cohes}$ & eV           & -5.006 & -4.906  & -4.999 & -4.902 & -5.92 & n/a   \\
 $T_{melt}$ & K             & $1850 \pm 50$ & $1800 \pm 50$  & $3000 \pm 100$  & $3200 \pm 100$ & n/a & 1650$^g$  \\
 $B$ & GPa                  & 257.4 & 267.0 &  &  & 223-226$^h$ &  222$^i$ \\
$c_{11}$ & GPa              & 513.1 & 399.2 & 513.6 & 537.2 &  & 420$^i$, 364.0$^j$ \\
$c_{12}$ & GPa              & 129.5 & 201.0 & 125.5 & 129.9 &  & 210, 92.0 \\
$c_{44}$ & GPa              & 43.3  & 73.7  & 45.1  &  52.0 &  & 95,  80.0 \\

\hline
  & & & & & & & \\
$\Delta H_{B20-B2}$ & eV/atom          & 0.128 & -0.13 & & & & $\sim -0.25^k$ & \\
  & & & & & & & \\
%    \hline\hline
%    \multicolumn{9}{l}{
%$^a$ denotes data which is not available from the literature or
%we could not derive by DFT calculations (e.g. $T_m$),
%}
%\bottomrule
\end{tabular}
%\end{ruledtabular}
%\footnotetext[3]{
%\footnotemark*{
%\footnotetext*{\raggedleft
\footnotetext*{
 denotes data which is not available from the literature or
we could not derive by DFT calculations (e.g. $T_m$),
DFT results obtained in the present work using the SIESTA package \cite{SIESTA}.
Details of the calculations are given in Table III. 
The net charges on Si: $n_q=-0.25e$ and $n_q=-0.21e$ for the B20 and B2 polymorphs, respectively.
The formation energy of $Fe_nSi_m$ is $\Delta H=\frac{1}{n+m} \biggm( E_{tot}^{n,m}-n E_{c,Fe}-m E_{c,Si} \biggm)$, where $E_{tot}^{n,m}$ is the calculated total free energy
% NSP: $E_{c,Fe}=-5.067$ and $E_{c,Si}=-4.536$ are the corresponding
of the $Fe_nSi_m$ system and $E_{c,Fe}=-5.933$ and $E_{c,Si}=-4.5495$ are the corresponding
atomic cohesive energies in their bcc and diamond phases as obtained by
the spinpolarized PBE DFT calculations.
If we take the mean experimental $\Delta H \approx 0.5$ eV/atom, hence
the estimated ("experimental") cohesive energy per atom of the B20 phase is
$E_c \approx (2 \times 0.5-4.63-4.28)/2=-4.995$ eV/atom for FeSi.
The experimental cohesive energies of Si (-4.63) and Fe (-4.28) have been used, respectively.
$\Delta H_{B20-B2}$: the enthalpy difference in eV/atom between the B20 and B2 phases at ambient conditions \cite{enthalpy},
$\gamma$ is the surface energy (eV/atom), $\gamma = E_{c,s} - E_{c,b}$, where
$E_{c,s}$ and $E_{c,b}$ are the average cohesive energy on the surface and in the
bulk, respectively.
$^b$: The experimental heat of formation is taken from ref. \cite{thermo_tmsil},
$^c$: A different melting temperature is given in ref. \cite{geology}, 
$^e$ refs. \cite{Petrova}, 
$^f$ ref. \cite{Miglio}, 
$^g$ \cite{meltB2}, 
$^h$: refs. \cite{Hafner} and \cite{Vocadlo}, 
$^i$, 
$^j$, 
$^k$ ref. \cite{enthalpy}, 
$^m$: For B2 FeSi $n_q=-0.1e$,
}
%\label{T4}
\end{ruledtabular}
\end{table*}
%\end{table}

%\begin{minipage}{6cm}
%\begin{tabular}{|l|l|}
%\hline
%1,2 & 1,2\footnote{This is a footnote.}
%                        \\\hline
%2,1 & 2,2               \\\hline
%3,1 & 3,2               \\\hline
%\end{tabular}
%\end{minipage}

%%%%%%%%%%%%%%%%%%%%%%%%
% T6
%%%%%%%%%%%%%%%%%%%%%%%
%\begin{table*}[tb]
\begin{table}[t]
\raggedright
%\sidecaption
\caption[]
{
The summary of the
basic results obtained by various long range electrostatics for the B20 and B2 polymorphs of FeSi
using the BOP-III+C force field.
}
%\begin{ruledtabular}
\begin{tabular}{cccccc}
\hline \hline
 B20 & Coulomb & $E_{cohes}$ (eV) & ${a_{latt}}$ ($\hbox{\AA}$) & $T_{melt}$ (K) & $B$ (GPa)  \\
\hline
% BOP   &           & & 4.475 & $1200 \pm 100$ & 217  \\
 BOP+C & DSFC      & -4.913 & 4.468  & $1400 \pm 50$ & 227   \\
       & Ewald     & -4.902 & 4.465 &  $1200 \pm 50$ & n/a   \\
       & PPPM      & -4.902 & 4.465 &  $1200 \pm 50$ & n/a    \\
       & Wolf      & -4.782 & 4.487 &   & n/a    \\
 Exp   &           &        & 4.49  &  $1450 \pm 50$  & <209  \\
%DFT   &           & -5.928 & 4.467 &  & 180  \\ \hline
\hline
 B2    &           &   &   &  &  \\ 
\hline
BOP+C  & DSFC      & -5.007 & 2.802 &  $1200 \pm 50$ & \\
       & Ewald     & -4.885 & 2.800 &  $1100 \pm 50$  & \\
       & PPPM      & -4.899 & 2.801 &  n/a          & \\
 Exp   &           &        & 2.83  &  $1650 \pm 50$ & 222  \\
% DFT   &           & -5.920 & 2.751 &  &  \\ \hline
\hline \hline
\end{tabular}
%\end{ruledtabular}
%\footnotetext[1]{
\footnotemark{
%The net charges of $q = \pm 0.25$ and $q =\pm 0.3$, 
The net charges of $q = \pm 0.15$ 
have been used for the B2 and B20 phases.
DSFC: damped shift forced coulomb \cite{Fennel}, Ewald-sum, PPPM: Particle-Particle Particle-Mesh 
using an iterative Poisson solver (details are given in the
corresponding LAMMPS manual).
The long range cutoff distance is $r_{cut}=12.0$ $\hbox{\AA}$.
Wolf: Wolf's summation method \cite{wolf} as implementes in LAMMPS.
%For Ewald and PPPM the Bulk modulus could not be calculated
%since
}
\label{T6}
%\end{ruledtabular}
%\end{table*}
\end{table}

%%%%%%%%%%%%%%%%%%%%%%%%
% T4
%%%%%%%%%%%%%%%%%%%%%%%%%
\begin{table*}[tb]nd
%\begin{table}[t]
%\sidecaption
\caption[]
{
The summary of the
basic results obtained for various polymorphs of FeSi
using BOP-I.
The DSFC method has been used for BOP+C.
}
\begin{ruledtabular}
\raggedright
\begin{tabular}{cccccccc}
%\hline
 polymorph & group & method  & $E_{cohes}$ (pw) & $E_{cohes}$ (DFT) & ${a_{latt}}$ (pw) & ${a_{latt}}$ (DFT) & ${a_{latt}}$ (EXP) \\
\hline
 FeSi & (B1, NaCl) &           &  & &   & &     \\
 & & BOP           & -4.909 & -5.181 & 4.410 & 4.95 &     \\
 & & BOP+C   & -4.817 & & 4.468 &  &    \\
 FeSi & (B3, ZnS) &           &  & &   & &     \\
%bopc: -4.311943   -89.311406    4.6587173
%bop 0.064684441   -4.5931855    8.5531515    4.6065427    3.6384177
 & & BOP             &  -4.593 & -4.97 & 4.607 & 5.26 &     \\
 & & BOP+C           & -4.312 &  & 4.659 &  &     \\
 Fe$_3$Si & (Cu$_3$Au)  &  & & & &  \\
 & & BOP           &  & -5.99 &  & 3.595 &      \\
 & & BOP+C           &  &  &  &  &      \\
 FeSi$_3$ & (CuAu$_3$)  &  & & & &  \\
 & &  BOP           &  & -4.75 &  & 3.64 &      \\
 & & BOP+C           &  &  &  &  &      \\
 $\alpha$-FeSi$_2$ & (tP3)     &  &   & & & &  \\
 & & BOP                      &   & -5.2 & a=2.685$^a$  & 2.66$^b$ & 2.69$^b$     \\
 &     &                     &  &  & c=5.072  & 5.08 & 5.134     \\
      & &  &   & & & &  \\
 &            &  BOP+C         &  &  &   & &    \\
 &                 &     &  &  & 5.135 &  &   \\
 $\beta$-FeSi$_2$ & (oC48)  &  &   & & &   \\
%9.8883394    7.9042551    7.8110161
 & & BOP                      & -4.376 & -5.571 & a=9.888 & 9.697$^b$ & 9.865$^c$    \\
 &  &                      &  &  & b=7.904 & 7.759 & 7.791$^c$    \\
 &  &                    &  &  & c=7.811 & 7.839 & 7.833$^c$    \\
    &  &  &   & & & &  \\
%-4.1729618   -235.60157    9.9569904    7.9591314    7.8652451 
 & & BOP+C        & -4.172 & -5.571 & 9.957 & &   \\
 &         &      &  & & 7.959 & &   \\
 &         &      &  & & 7.865 & &   \\
 $\gamma$-FeSi$_2$ & (C1, CaF$_2$)  &  &   & & &   \\
  &  & BOP & -3.950 &  & 4.416 &  &   \\
  &  & BOP+C & -3.950 &  & 4.416 &  &   \\
 FeSi$_3$ & (DO$_3, AlFe_3$) & BOP-V & -4.129 & -5.259 & 4.232 & 5.117 &   \\
  &  & BOP+C & -4.080 &  & 4.244 &  &   \\
 Fe$_3$Si & (DO$_3, AlFe_3$) & BOP-V & -4.239 &  & 3.979 &  &   \\
  &  & BOP+C & -3.885 &  & 4.294 &  &   \\
\end{tabular}
%\end{ruledtabular}
%\footnotetext[1]{
\footnotemark{
'pw' denotes present work,
ref. \cite{Hafner},
$^b$ present work obtained by SIESTA using relaxed cell calculation, $^c$ ref. \cite{betafesi2},
The net charges of $q = \pm 0.15$ were used for B1.
%The net charges of $q = \pm 0.21$ and $q =\pm 0.25$, 
%have been used, respectively for the B2 and B20 phases.
DSFC: damped shift forced coulomb \cite{Fennel}.
%DSFC: damped shift forced coulomb \cite{Fennel}, Ewald-sum, PPPM: Particle-Particle Particle-Mesh 
%using an iterative Poisson solver (details are given in the
%corresponding LAMMPS manual).
The long range cutoff distance $r_{cut}=12.0$ $\hbox{\AA}$.
}
\label{T7}
\end{ruledtabular}
\end{table*}
%\end{table}

%%%%%%%%%%%%%%%%%%%%%%%%
% T8
%%%%%%%%%%%%%%%%%%%%%%%%
%\begin{table*}[tb]
\begin{table}[t]
%\sidecaption
\caption[]
{
The summary of elastic properties and bulk modulus (GPa) 
obtained for the B20, B2 and $\beta-FeSi_2$ polymorphs of FeSi.
}
\begin{ruledtabular}
\begin{tabular}{ccccc}
\hline
  & BOP & BOP+C & dft & exp  \\
\hline
B20 ($\epsilon$-FeSi) & & & & \\
\hline
C11 & 363.3 & 363.3 & 488$^a$ & 316.7$^b$ \\
C12 & 161.4 & 149.6 & 213 & 112.9 \\
C44 & 89.5 & 83.4 & 125 & 125.2  \\
B$^c$   & 228.7  & 230.4 & 176.2 & 180.8 \\ 
Y$^d$   & 264.1 & 277.1 & 358.6 & 257.4 \\
$\mu^e$ & 0.31 & & 0.26 & \\
\hline
B2 (CsCl) & & & & \\
\hline
C11 & 414.1 & 318.4 &  & \\
C12 & 186.9 & 239.0 &  & \\
C44 & 101.1 & 166.5 &  & \\
B$^c$   & 262.6  & 265.5 & 223$^f$ & 222$^f$  \\ 
Y$^d$   & 297.9 & 113.445 &  &  \\
\hline
$\alpha-FeSi_2$ (oC48) & & & & \\
\hline
C11 &  & 120.5 &  &  \\
C12 &  & 47.3 &  &  \\
C44 &  & -23.8 &  &  \\
B$^c$   &  &  & 172$^f$ &   \\ 
\hline
$\beta-FeSi_2$ (oC48) & & & & \\
\hline
C11 & 267.6 & 266.8 & 314.6$^g$ & 264$^h$ \\
C12 & 60.3 & 133.4 & 95.8 & 177$^h$ \\ 
C22 & 187.2 & 381.7 & 355.8 & 78$^h$ \\ 
C23 & 43.8 & 138.2 & 89.4 & \\ 
C33 & 185.2 & 252.2 & 362.5 & \\
C44 & 53.3 & 109.5 & 126.4 & \\ 
C55 & 38.7 & 36.3 & 127.0 & \\ 
C66 & 38.8 & 32.0 & 142.0 & \\ 
B$^c$   & 129.4 & 177.9 & 172.5 & 206$^h$  \\ 
Y$^d$   & 245.4 & 177.9 & 312.0 &  \\
$\mu^e$ & 0.184 & 0.333 & 0.199 & \\
%\hline
%C14 & -1.720223653e-08 & 1.782052436e-08 & & \\
%C15 & -2.450252242e-08 &
%C16 & -1.782004273e-08 &
%C24 & 1.082863395e-08  &
%C25 & -7.796259295e-09 &
%C26 &  2.157913625e-09 &
%C34 & -7.757089775e-09 &

%Elastic Constant C35all = -1.141595302e-08 GPa
%Elastic Constant C36all = -5.84718556e-09 GPa
%Elastic Constant C45all = 2.060875312e-08 GPa
%Elastic Constant C46all = -1.464409184e-08 GPa
\end{tabular}
\end{ruledtabular}
%\footnotetext[1]{
\footnotemark{
ref. \cite{Vocadlo},
$^b$ ref. \cite{Petrova},
$^c$ $B = \frac{1}{3} (c_{11}+2 c_{12})$ for anisotropic
cubic crystals.
$^f$ ref. \cite{Hafner}, $^g$ ref. \cite{Tani}, $^h$ ref. \cite{Miglio}
For the Youngs's modulus and for the Posisson ratio the 
relations
$Y=\biggm[ (c_{11}+2c_{12}) \frac{c_{11}-c_{12}}{c_{11}+c_{12}} \biggm]$
and
$\mu=\frac{c_{12}}{c_{11}+c_{12}}$
have been used.
}
\label{T5}
%\end{ruledtabular}
\end{table}

%\vspace{-0.7cm}

\section{Results}

\subsection{Melting properties}

The melting simulations have been run under NPT conditions
until 1 ns using Nose-Hoover thermostat (and prestostat) as implemented
in the LAMMPS code \cite{parcas}.
The mean square of atomic displacements and the rdf file have been
analysed and compared together with the visual inspection
of movie (animation) files which provide sufficient information
on melting.
The sharp melting transition can be seen in Fig. 3 together
with the radical change in the rdf (Inset Fig. 4).

 As mentioned earlier, each variants of BOP-II give rather large melting point of
$T_m \approx 2500 \pm 100$ K which is attributed to the
overbinding of the potential at short interatomic distances.
This is the typical consequence of the $D_0/r_0 < 3.0$ ratio.
Using BOP-I $T_m$ has been lowered effectively, especially with
the BOP+C variant which gives $T_m$ nearly in perfect agreement
with experiment.
BOP-III underestimates $T_m$ by $150$ K due to the too large
$D_0/r_0$ ratio.

 In order to get more insight into the adjustment of $T_m$ we
have studied the variation of it as a function of the dimer parameters $D_0$
and $r_0$.
In Fig. 3 one see that $T_m$ is nearly not sensitive to
$D_0/r_0$ within the regime of natural values ($D_0 < 4$ eV/atom,
$r_0 > 1.8$ $\hbox{\AA}$) and remains at around $T_m \approx 2200 \pm 200$ K. 
However, when $D_0/r_0 > 4.0$ 
$T_m$ starts to drop.
This is not surprising, if we consider that
in this way we introduce a structural anisotropy into the system.
The BOP tries to shorten first neighbor distances to the
unphysical $r_0 < 1.6$ $\hbox{\AA}$, which is hindered, however,
by the repulsive potential in the solid environment put by the
first neighbors on a central atom.
The reason of the sensitivity of $T_m$ to $D_0/r_0$ is somewhat unclear.
%Therefore, finally we end up with correct first neighbor distances 
%and lattice constant if $D_0/r_0$ is properly set.
Nevertheless, the stress what is put by the short $r_0$
which tries to shorten the first neighbor distances
reduces $T_m$ effectively
when compared with BOP-II with natural
dimer properties ($r_0 \approx 2.1$ $\hbox{\AA}$, $D_0 \approx 3.0$ eV/atom).
Since in the solid state the first neighbor distance can not be shorten seriously due to
the strong repulsion of the first neighbors the lattice constant
remains nearly unaffected to those cases where
$D_0/r_0 < 3.0$ (natural dimer regime).
Hence, introducing an artificial anisotropy into the radial parameter set
pushes $T_m$ in the right direction.
Using then an appropriately chosen $D_0;r_0$ pair
one can tune $T_m$ effectively.
The price what we pay is that the dimer is incorrectly described with
this parameter set. However, our aim is to develop a BOP which
describes correctly solid state with various polymorphs and not molecules.
Hence, the drop of the constraint put by the Pauling relation
on the initial guess parameters could lead to more effective BOP force fields
in the soild state.
At least the overall performance of the BOP-III set as it has been
shown in Table (\ref{T4})
does not show serious change in the test results.

\subsection{Phase order problem: BOP+C}

  BOP fails to reproduce correctly the energetic relationship (phase order) between
the two most stable polymorphs (B20 and B2) of FeSi and the
B2 phase is more stable by $\sim 0.1$ eV (see Table 3).
Table 2 reports us that 
ab initio calculations slightly favors B20 over B2, and the energetic difference
is rather small, less than $0.01$ eV, which is within the accuracy of 
present day exchange-correlation functionals for cohesive energy.
Although the difference is not negligible between the phases,
one can conclude that it should be accounted for by any reliable approach (empirical,
or ab initio).
Therefore we decided to improve the performance of the BOP emprirical potential
by adding a simple Coulomb term to it as it has been shown in Eq (~\ref{bop+c}).

  However, it is rather difficult to give reliable net atomic charges $q_i$
required by the Coulomb term. As it is well known 
Mulliken charges are exaggerated. Instead we use Bader charges based on the
atom in molecule framework \cite{Bader}.
The obtained partial charges for Si and Fe ($\pm 0.15$) both in B20 and B2
FeSi are shown in Table II.
The net charges are calculated using the self-consistent density files
obtained by SIESTA \cite{SIESTA}.
Note that the net Mulliken charges and those obtained
by the Bader's are rather different. The Mulliken's approach
provides larger net charges in B20 while Bader estimates smaller $q$ in
B20.
What is for sure that the Mulliken charges are too large, however, it is
far not trivial whether the Bader AIM charges are in the correct order.
One has to keep in mind that the decomposition of the space in the system
could result in small variations in the magnitude of $q$ which
could reverse the order of the charges system by system.
Hence we take with some caution the obtained charges, especially
their relationship to each other.

 Therefore, in this way, we get still the B2 phase more stable since
SIESTA provided somewhat larger $q$ for B2.
We also observe, however, that the relative stability of B2 FeSi
is lost when the net charges are slightly increased to $\pm 0.23$ in the B20 phase.
Also, the net difference of $\Delta q < 0.05$ between the net charges of the B2 and B20
phases (e.g. $q_{B2} \approx 0.21$ and $q_{B20} \approx 0.25$)
will slightly stabilize the B20 phase by $0.015$ eV/atom.
In other words, the problem is slightly sensitive to the choice of $q_i$.
Unfortunately the choice of the appropriate $q_i$ is somewhat arbitrary,
since the obtained Bader charges seems to be understimated while the
Mulliken charges are overtestimated.
We find in other compounds, such as ZnO e.g., that somewhat a larger
point charge is required than provided by SIESTA by $\sim 10-20 \%$ to obtain
consistent properties with experimental and ab initio ones using
e.g. Buckingham and Coulomb potential. \footnotetext[2]{Unpublished results (2011)}
 
 Moreover, the charge distribution as being nonlocal, the point charge
model can only be relaible if the net charge is properly adjusted.
There is no standard way of setting in point charges in Coulombic models
and often it can be done only on a trial and error basis.
We also find that the choice of $q \approx 0.25e$ in B20 FeSi and
$q \approx 0.15e$ in the B2 phase gives satisfactory results.
In this way the B20 polymorph becomes slightly more stable.
Hence the proper choice of the net charges around the value obtained
by {\em ab initio} calculations could account for the required
stability relationship between the various polymorphs.
Nevertheless, it would be fruitfull to look for a more adequate
physical modell which can capture the main essence of
the polymorphs without the adjustment of net charges.
A possible choice could be the fitting of the COMB modell to FeSi \cite{comb}.
This modell is based on the Tersoff BOP modell, and the long range
part should only be adjusted.
This definitly goes beyond the scope of the present paper.

\subsection{Results on the B20 $\rightarrow$ B2 phase transformation}

\subsection{Results on disilicides}
Iron disilicide exists in two stable modifications, the room temperature phase $\beta$-FeSi$_2$ which is orthorhombic and the high temperature phase $\alpha$-FeSi2 which is tetragonal. Both phases show very interesting properties for potential applications in thermoelectrics, photovoltaics and optoelectronics
\cite{disilicides}.

\subsection{Si-$\beta$-FeSi$_2$ heterostructures: simulated annealing}

\section{acknowledgement}
%{\scriptsize
%This work is supported by the OTKA grant
% K-68312
%from the Hungarian Academy of Sciences.
%Support from the bilateral German-Hungarian
%exchange program DAAD-M\"OB (Grant No. 37-3/2008)
%and German Science Foundation (DFG research
%group 845, project HE2137/4-1) is also
%acknowledged.
The kind help of P. Erhart (Darmstadt) is greatly acknowledged.
We wish to thank to K. Nordlund (Helsinki) 
for helpful discussions and constant help in the use
of the PARCAS code and to S. Plimpton (Sandia) in the use of the
LAMMPS code.
The calculations (simulations) have been
done mostly on the supercomputers
of the NIIF center (Budapest).
The help of the staff is also greatly appreciated, 
mostly the constant help of P. Stefan.
%The work has been performed partly under the projects
%HPC-EUROPA and PRACE with the support of
%the European Community using the supercomputing
%facility at CSC in Espoo and at GENCI/CEA (France).

%\vspace{-0.7cm}

\end{document}